# Magnetization of 2.6T in gadolinium thin films

G. Scheunert*, W. R. Hendren, C. Ward, and R. M. Bowman[#]

Centre for Nanostructured Media, School of Mathematics and Physics, Queen's University Belfast, Belfast BT7 1NN UK

## ABSTRACT

There is renewed interest in rare-earth elements and gadolinium in particular for a range of studies in coupling physics and applications. However, it is still apparent that synthesis impacts understanding of the intrinsic magnetic properties of thin gadolinium films, particularly for thicknesses of topicality. We report studies on 50nm thick nanogranular polycrystalline gadolinium thin films on $SiO_2$ wafers that demonstrate single-crystal like behavior. The maximum in-plane saturation magnetization at 4K was found to be $4\pi M_S^{4K} = (2.61\pm0.26)$T with a coercivity of $H_C^{4K} = (160\pm5)$Oe. A maximum Curie point of $T_C = (293\pm2)$K was measured via zero-field-cooled - field-cooled magnetization measurements in close agreement with values reported in bulk single crystals. Our measurements revealed magnetic transitions at $T_1 = (12\pm2)$K (as deposited samples) and $T_2 = (22\pm2)$K (depositions on heated substrates) possibly arising from the interaction of paramagnetic fcc grains with their ferromagnetic hcp counterparts.

Corresponding authors: * gscheunert01@qub.ac.uk   [#] r.m.bowman@qub.ac.uk

Gd has been recognized as a promising candidate for high magnetic moment applications at cryogenic temperatures up to room temperature [1][2]. Among the rare earth metals it has notable attributes such as a high Curie point of 293K and a saturation magnetization of 2.66T (in single-crystals [3]). In the 1990's Gd-W systems were studied extensively while investigating surface enhanced magnetism [4] and there is now renewed interest in Gd thin-film properties particularly with regards to fundamental studies on the possibility of ferromagnetic - normal metal - antiferromagnetic coupling to exploit the 4f shell with 7.6 $\mu_B$ magnetic moment per atom to increase the moment of 3D ferromagnets such as in Fe/Cr/Gd [5] and in exchange interactions in permalloy/Gd [6] as well as Fe/Gd/Fe systems [7]. Recent advances in the field of nano- and microstructure cooling [8][9] along with the possibility of rare earths featuring in contemporary applications such as high magnetic moment pole heads for hard drives [10] also provide applied context. To date it is also noticeable that the fundamental experimental studies have been hampered by the fact that the magnetization values for thin films of Gd [11] have always been significantly smaller than seen in bulk and single-crystal samples [3]. This has been attributed to various causes such as the granular structure and resulting finite size effects as well as shape related reasons due to the transition from 3D to 2D magnetic systems in thin films [12]. Altogether these factors necessitate and encourage renewed interest in optimised Gd thin films and the physics they present.

In this letter we report the creation Gd thin films with magnetic response akin to single crystals along with the observation of transitions that suggest strong intergranular interactions. Further, from the applied perspective, this has been achieved on silicon oxide by the scalable process of sputtering.

From the outset, oxidation in rare earth films is an obvious obstacle to obtaining fundamental properties and so to avoid this in Gd thin films impacting on the magnetic properties [13] various approaches have been considered such as using thicker layers of around 700nm and more [11] as well as tungsten multilayer stacks [14][15] and aluminum, titanium, or silicon capping [16]. We established that using 5nm Ta seed and capping layers provided effective protection in our systems. All films were prepared by DC magnetron sputtering on Si wafers with a 200 nm thermal oxide under ultra-high vacuum conditions (base pressure $< 3 \times 10^{-9}$ mbar) using research grade argon process gas (process pressure $10^{-3}$ mbar) and a 99.8% pure Gd target. The sputtering targets were placed a distance of 16cm, at angles of incidence of 25° for Gd and 15° for Ta. Deposition rates were determined by quartz crystal and film thickness was confirmed by XRD (X-ray diffraction) and TEM (transition electron microscopy). Prior to deposition, oxygen contamination was mitigated by outgassing the sample holder and wafer at the process temperature of 350°C while monitoring the vacuum using a residual gas analyser. Targets were then pre-sputtered for 30 min at 15 W/in$^2$. Traces of oxygen are readily identified via the magnetic properties of the samples, i.e. reduced spontaneous magnetization (oxygen quantities of a few atomic per cent already reduced the saturation magnetization by 20%). The purity of films was additionally verified by Auger spectroscopy which provided clear evidence for the effectiveness of Ta as a barrier material, preventing post-deposition corrosion under standard atmospheric conditions as well as oxygen diffusion from the wafer's oxide layer. Samples showed identical magnetization values even after months of storage under normal atmospheric conditions.

Crystallographic analysis revealed a polycrystalline nanogranular structure of the Gd films. The stacks of Ta(5nm)/Gd(50nm)/Ta(5nm) on SiO$_2$ wafers showed a very clean interface at the seed layer and a hilly appearance at the capping side which is due to increased grain formation towards the top as shown in Figure 1(a). For samples prepared on heated wafers at 350°C we observe hcp (hexagonal close-packed) structure with a = (3.60±0.02)Å and c = (5.78±0.02)Å for the majority of the grains as reported for single-crystal bulk Gd [17], where the (0 0 2) peak dominated XRD scans as shown in Figure 1 (b) trace (B) and (C). Samples prepared at lower deposition rates (0.40Å/s) is dominated by only the hcp (0 0 2) peak, trace (B), hence the c-axis of grains predominately points out-of-plane, which is the magnetic easy-axis in single crystals. For samples deposited at room temperature all Gd related peaks were shifted by 0.3° towards smaller angles resulting in differing hexagonal lattice parameters a = (3.36±0.02)Å and c = (5.61±0.02)Å, which indicates interfacial and internal strain of those layers. A second metastable fcc (face-centred cubic) phase was found with a = (5.33±0.02)Å as reported previously for Gd thin films [18], nanoparticles [19][20][21], and nanostructures [22]. This fcc phase occupied a significant portion of the thin films deposited at room temperature according to Θ-2Θ scans, Figure 1 (a) trace (A), but was not as visible in gracing incidence scans. This suggests the fcc phase predominantly exists at the seed layer interface, probably due to stacking faults [23][24], and in much smaller quantities throughout the film due to size-induced structural phase transformation [22] as reported for grains of ~10nm and disappears with further nucleation towards the capping layer. XRD scans on films deposited on a heated substrate (350°C) showed small signs of this fcc phase in similar proportions in Θ-2Θ and gracing incidence scans, suggesting a small overall content and a generally much more homogeneous texture of the films [25]. The Scherrer method for calculating the average grain size [20][26] was applied to the hcp (0 0 2) peak. As only the generic background correction in the XRD was used the absolute values have a considerable error margin, however the relative size relationship is not affected as the same systematic errors apply to all samples. For heated depositions the average grain size (~ 30nm) is larger than for room temperature depositions (~ 20nm) and for both substrate temperatures there is a trend towards larger grain sizes for increased deposition rates, e.g. increasing the rate from 1.23Å/s to 3.10 Å/s suggests ~2nm larger grains.

Magnetic measurements were conducted with a commercial Quantum Design MPMS (magnetic property measurement system) SQUID (superconducting quantum interference device) in fields up to 50kOe (magnetic moment uncertainty of ~5% for values as small as $10^{-5}$ emu). Samples were cut into pieces of ~5x5mm$^2$ with the exact area measured via digital image capture and then mounted in transparent drinking straws to reduce background noise [27]. With an emphasis on establishing saturation values in-plane measurements of the spontaneous magnetization at H = 50kOe were carried out. Although Gd is still not fully saturated at 50kOe previous studies showed further increase in the H field does not lead to a significant increase in the magnetization [3] therefore we continue to use the terminology saturation magnetization for these measurements described.

In agreement to Φ-angle independent XRD patterns magnetization values are similar for all in-plane directions. This is due to a random in-plane orientation of the nanosize Gd grains which reduces the very small crystal anisotropy of Gd, hence, shape anisotropy dominates. Comparing in-plane saturation measurements with literature for single crystals of bulk Gd [3][28] we find qualitative agreement of the M-T curves as well as similar Curie temperatures of $T_C$ = (293±2)K for depositions on heated substrates (350°C) and easy-axis (c-axis) measurements. Absolute saturation values of up to $4\pi M_S^{4K}$ = (2.61±0.26)T at 4K, achieved at a deposition rate of 1.23Å/s & 350°C substrate temperature, almost match single-crystal easy-axis values. The remaining small discrepancy probably arises from a very small crystal anisotropy contribution as well as interfacial (grain boundary) magnetic anisotropies such as reported before for nanogranular Gd synthesised via inert-gas condensation [29]. The saturation magnetization slightly drops again for higher deposition rates (3.10Å/s) which might be due to small structural differences, although XRD patterns only revealed minimally strained lattice parameters. A more drastic drop of the saturation down to $4\pi M_S^{4K}$ = (2.19±0.22)T was found for low-deposition-rate samples (0.40Å/s). Those samples suffer from a stronger crystal anisotropy contribution, bringing down the in-plane moment, as the c-axis of nearly all grains points out-of-plane. The zero-field-cooled - field-cooled (ZFC-FC) measurements presented in Figure 2 (b) confirm this where the sample (0.40Å/s, 350°C) shows similar behavior to the b-axis of a single crystal, i.e. the magnetic hard-axis [28]. Complementary FC graphs of c-axis and b-axis measurements on Gd single crystals have led to the conclusion the magnetic easy axis effectively rotates, i.e. for a temperature range of ~240K down to ~80K not the c-axis but the b-axis becomes the magnetic easy axis. This effect was not observed in our heated high deposition rate samples (1.23Å/s & 3.10Å/s, 350°C), where the magnetic easy axis stays in-plane due to shape anisotropy dominating crystal anisotropy. Lowest saturation values of $4\pi M_S^{4K} = (2.05\pm0.21)$T were found for low deposition rate samples (0.40Å/s) deposited at room temperature, in which a high content of the fcc phase and smaller grain sizes lead to a significant contribution of intergrain and intragrain anisotropies [29].

A common feature of all samples is a kink in the ZFC-FC measurements at $T_1 = (12\pm2)$K for room temperature depositions and $T_2 = (22\pm2)$K for the high-rate heated depositions. There is still an ongoing debate in literature whether this feature, reported before for room temperature depositions on Mo seed layers [12], structurally very similar to Ta, is due to superparamagnetic Gd particles or an additional magnetic transition. An in-depth discussion is further complicated by observation of size-induced phase transformation of small grains from hcp to fcc at ~10nm which is also roughly the threshold for superparamagnetism [21] and in addition there has never been a separate investigation of the Gd fcc phase magnetic properties due to its metastable nature. Considering that the lattice parameters of the fcc phase are the same for all samples (room temperature and heated depositions) but its content changes dramatically with substrate temperature, explanation attempts should resort to the interaction of the fcc and hcp grains which the Gd thin films are comprised of [30]. Assuming different anisotropies for the two grain types qualitatively explains the observed magnetization behavior and phase transitions, but clarification can only be provided by magnetic modelling which is not the scope of this letter. As deposited samples had a lower Curie point of $T_C = (278\pm2)$K due to their smaller grain size, which was identified as the main reason for a drop in $T_C$ in previous studies [20]. Figure 3 shows hysteresis loops recorded at 4K, 20K, and 50K. Coercivity values match expectations derived from the grain sizes [12], heated depositions with larger grains (3.10Å/s, 350°C) lead to a smaller coercivity of $H_C^{4K}$ = (160±5)Oe while the room temperature depositions show three times larger coercivity, $H_C^{4K}$ = (495±5)Oe. Further hysteresis loops measured at 20K and 50K for those samples did show decreasing coercivity values $H_C^{20K}$ = (405±5)Oe and $H_C^{50K}$ = (330±5)Oe, whereas the large-grain samples' coercivity changes remained within the uncertainty range. Heated depositions sputtered at higher rates (3.10Å/s) showed a harder hysteresis loop than low rate ones (0.40Å/s) possibly due to slightly larger grains but also differences in their orientation. The hysteresis loops also suggest a degree of increasing coupling with Figure 3(a), at low temperature reminiscent to Stoner-Wolfarth behavior whilst the enhanced remanence in Figure 3(b) & 3(c) indicate probable exchange coupling which could be larger in Figure 3(c). Room temperature depositions showed small in-plane remnant fields at 4K of $4\pi M_{rem}^{4K}$ = 0.71T, whereas the 350°C deposited samples lead to fields of $4\pi M_{rem}^{4K}$ =0.91T for deposition rates of 0.40Å/s and $4\pi M_{rem}^{4K}$ =1.44T for highest deposition rate of 3.10Å/s, i.e. the remnant field scales with the grain size as larger grains have higher thermal stability. The remnant field dropped to zero magnetization when samples were heated up in zero field at $T_0$ = (250±5)K and $T_0$ = (275±5)K for room temperature and heated depositions, respectively.

In summary thin Gd layers, sandwiched by Ta, sputtered on $SiO_2$ wafers were found to be corrosion resistant under normal atmosphere conditions. By heating the substrate during the deposition to 350°C magnetic properties very close to Gd single crystals were achieved. Key data was a Curie point of $T_C$ = (293±2)K and a saturation magnetization at 4K of $4\pi M_S^{4K}$ = (2.61±0.26)T for films prepared at deposition rates of 1.23Å/s. We have identified the main reasons for reduced magnetic properties of samples deposited at room temperature to be smaller grain sizes, a larger proportion of a paramagnetic fcc Gd phase, and residual strain in the sputtered films. The thin Gd layers showed a magnetic transition that might arise from an interaction of paramagnetic fcc and ferromagnetic hcp particles at $T_1 = (12\pm2)$K and $T_2 = (22\pm2)$K for as deposited and heated depositions, respectively. This understanding and control will now facilitate future studies in rare-earth, and in particular Gd, ferromagnetically coupled thin-film systems.

## ACKNOWLEDGEMENTS

We are grateful to Roy Chantrell (University of York) and Thomas F. Ambrose for their helpful insights. We thank Seagate Technology (Ireland) for their financial support to establish ANSIN (www.ansin.eu).

**FIGURE CAPTIONS**

**Fig. 1 (colour online)** (a) TEM cross-sectional view of a Gd layer sputtered at 350°C on a Si wafer and (b) XRD spectra (Θ-2Θ and gracing incidence scans) of samples sputtered on a SiO$_2$ wafer at (A) room temperature & 0.40Å/s, (B) 350°C & 0.40Å/s, and (C) 350°C & 3.10Å/s.

**Fig. 2 (colour online)** (a) In-plane saturation magnetization (H = 50kOe) and (b) in-plane ZFC(→)-FC(←) moments (H = 100Oe) of pure Gd layers on SiO$_2$ wafers, revealing magnetic transitions. Both with respect to the sputtering parameters (deposition rate and substrate temperature).

**Fig. 3 (colour online)** In-plane hysteresis loops recorded at 4K, 20K, and 50K, for Gd layers sputtered at (a) room temperature & 3.10Å/s, (b) 350°C & 0.40Å/s, and (c) 350°C & 3.10Å/s.

**Fig.1**:

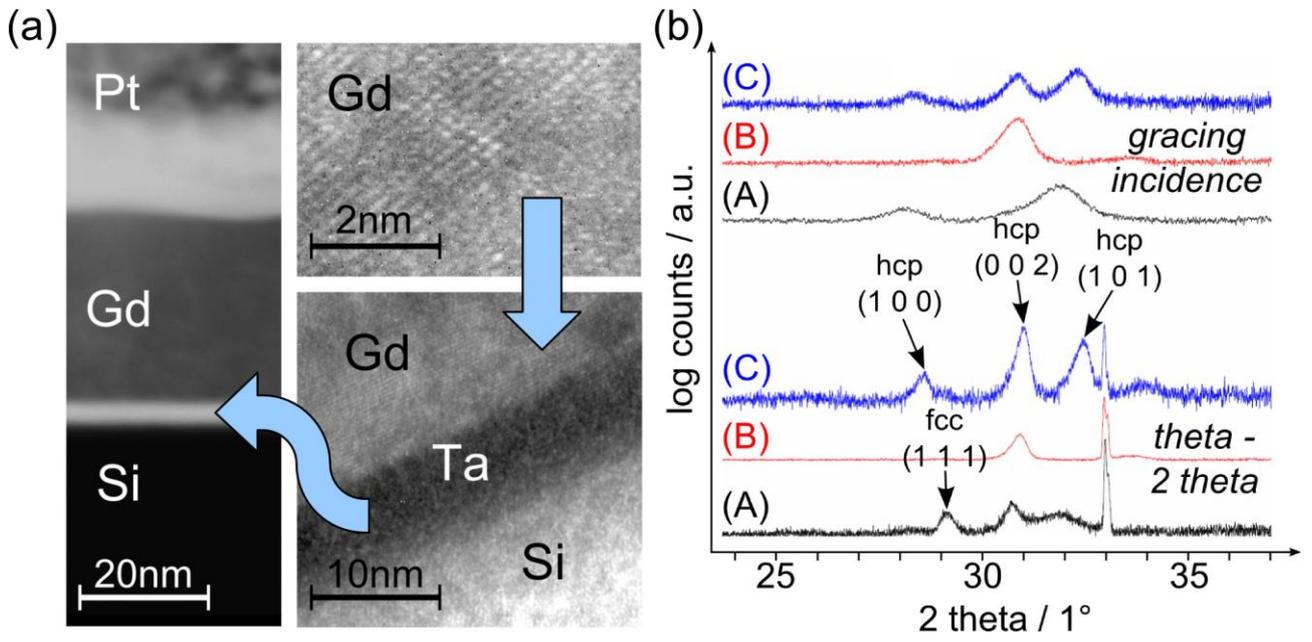

**Fig.2**:

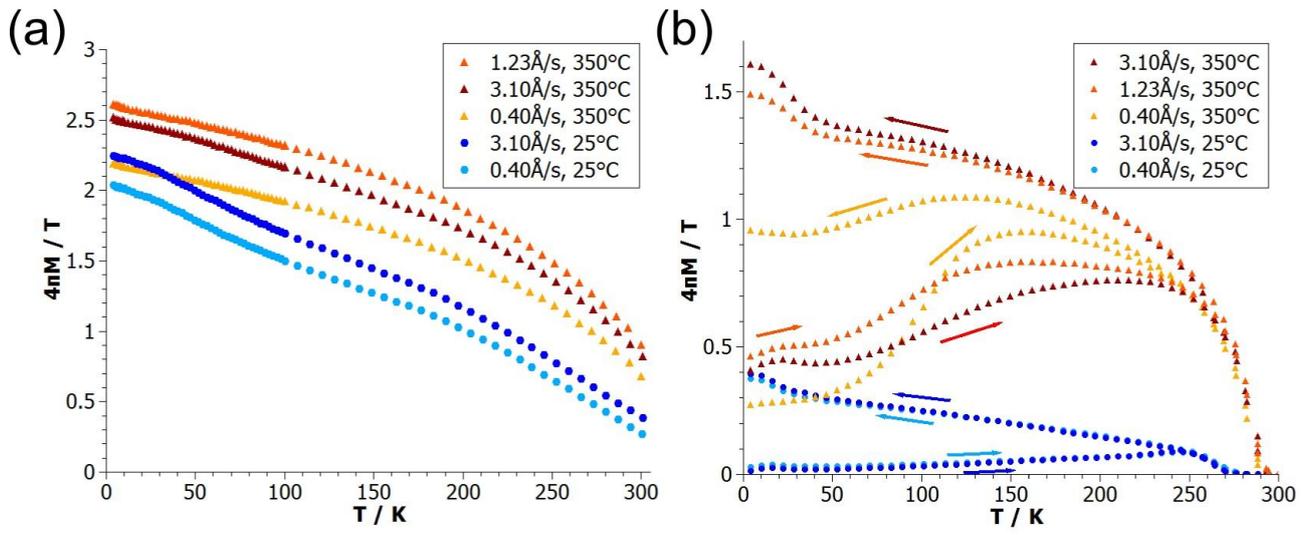

**Fig.3**:

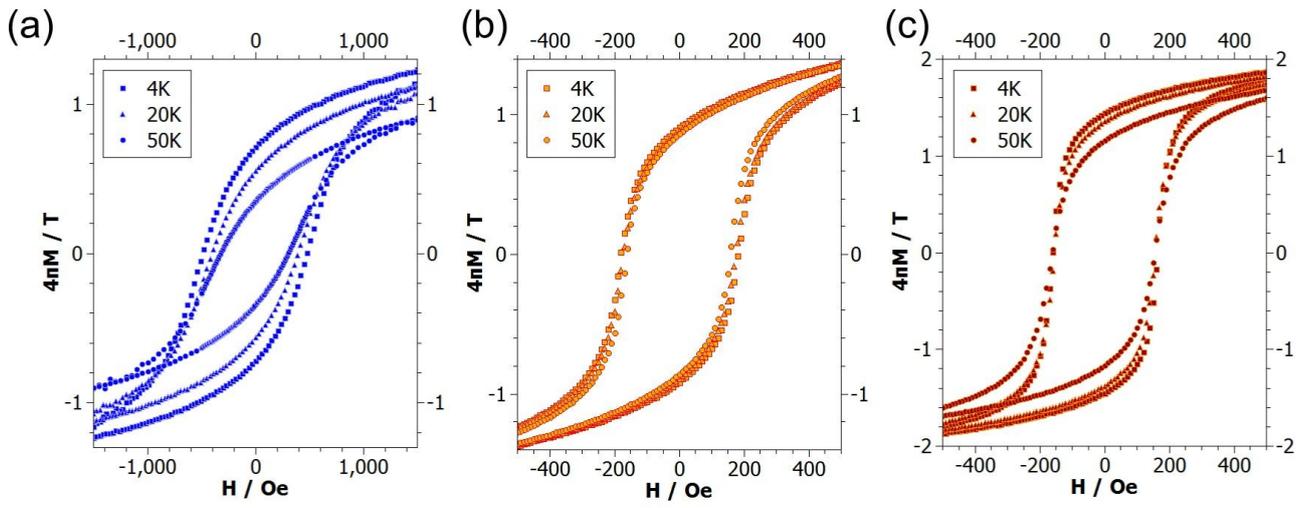